\documentclass[aps,prl,twocolumn,groupedaddress,showpacs]{revtex4-1}

\usepackage{graphics}
\usepackage{dcolumn}
\usepackage{bm}

\begin{document}


\title{The Anomalous Compressibility and Metallization of Deuterium under Shock-Wave Compression}


\author{A.L. Khomkin and A.S. Shumikhin}
\email{shum\underline{ }ac@mail.ru}
\affiliation{Joint Institute for High Temperature, Moscow 125412,
Russia}


\date{\today}

\begin{abstract}
High values of deuterium compressibility under shock-wave
compression, recently discovered in several experiments, are
explained by an unusual dielectric-metal phase transition from a
dense molecular gas into a liquid-metal atomic gas. This phase
transition recently described by authors was named the
dissociative phase transition (DPT). The same phase transition
describes a significant scattering of experimentally measured
densities within an area of anomalous compressibility.

\end{abstract}


\maketitle

\section{I. Introduction}

The experiments on liquid deuterium shock-wave compression were
widely discussed recently \cite{Fortov}. The first experiments
\cite{DaSilva} discovered high (even abnormally high)
compressibility, but their results had significant uncertainties.
The further experiments that used different methods did not reveal
such an anomalous compressibility \cite{Knudson,Nellis,Trunin}. A
possibility that the first experiments were not correct were
discussed, however the authors of \cite{DaSilva} did not denounce
their results. The discussion of experiments together with
numerous theoretical models can be found in the review
\cite{Fortov} and in the original papers. It is important to note
that all theoretical calculations gave a continuous density
dependence of pressure on Hugoniot curves in the experimentally
studied areas \cite{Knudson,Trunin, Nellis}. Only one rather
exotic model of "linear mix" by M. Ross \cite{Ross} features the
anomalous compressibility in agreement with the results of
\cite{DaSilva}. The deuterium free energy in this model is a
linear mixture of molar fractions of atomic-molecular and metallic
components. The Ross model describes the atomic-molecular
component in a usual way, within the frame of perturbation theory
of liquid mixture of atoms and molecules. The liquid metal
component is described in the approximation of amount of
degenerated electrons and classical ions. In this model, the shock
adiabatic curve is a continuous function of the volume (specific
density) like in all other models \cite{Fortov}, but the curve
demonstrates high compressibility values owing to dominance of the
liquid-metal component at high densities. Thus, the Ross model
explains certain correlation between the high values of
compressibility and deuterium metallization.

In \cite{Khomkin}, we suggested a new model for hydrogen molecule
dissociation related with metallization of an atomic component of
molecular hydrogen under compression.  The conversion of a dense
molecular liquid into an atomic metallized liquid is the
first-order phase transition, DPT. The model assumes that, owing
to the appearance of conductivity electrons (cohesion), the
interaction of free (dissociated) atoms in dense molecular
hydrogen turns into a collective interaction. Such interaction is
well known in the theory of liquid alkali metals, and the binding
energy is known as the cohesion energy. In \cite{Khomkin}, we
calculated the cohesion energy for all densities in the frame of
Wigner-Seitz-Bardeen theory. The collective binding energy grows
comparable with the binding energy of an atom in a molecule (half
of the dissociation energy) thus increasing the dissociation rate.
The estimate of the critical transition point made in
\cite{Khomkin} indicates that the transition occurs in the anomaly
area of Hugoniot curve for deuterium.

In the present work, we calculate the shock-wave Hugoniot curve
for deuterium in the low-temperature branch (T $\alt$ 12000\,K) on
the basis of the model \cite{Khomkin}. The calculated Hugoniot
curve demonstrates anomalous behavior: the Rankine-Hugoniot
equation has one, two, three, and again two and one solutions in
the certain pressure interval (in the anomaly area).  The third
solution is unstable and corresponds to the unstable part of the
curve. In this case, one can consider a break of the adiabatic
curve in the area of the anomaly. This area coincides with the
area of experimental data with high uncertainties
\cite{DaSilva,Knudson,Nellis,Trunin}. Therefore, there is a
correlation of the anomaly area with the two-phase state on the
adiabatic curve, where the molecular fluid (gas with a density of
a liquid) and atomic metallized liquid coexist. In this area, the
density is not a well-defined parameter, which results in the
scattering of experimental data on density values.

\section{II. Chemical model for atomic-molecular deuterium}

The free energy of dissociated atomic-molecular mixture of $N_{m}$
molecules and $N_{a}$ atoms in volume $V$ at temperature $T$ in
the liquid perturbation theory suggested in \cite{Khomkin} has the
following form:

\begin{eqnarray}
 F = -N_{a}kT \ln(\frac{eVg_{a}}{N_{a}\lambda_{a}^{3}})
 -N_{m}kT \ln(\frac{eVg_{m}\Sigma_{m}}{N_{m}\lambda_{m}^{3}})\nonumber \\
 +(N_{a}+N_{m})kT\frac{4\eta-3\eta^{2}}{(1-\eta)^{2}} + \frac{1}{2}N_{a}E_{coh}(y).
\end{eqnarray}

where $\lambda_{a(m)}=(2\pi\hbar^{2}/m_{a(m)}kT)^{1/2}$ and
$g_{a(m)}$ are the de Broglie wave lengths of an atom (a molecule)
and their statistical weights, respectively, $\Sigma_{m}$ is the
partition function of a molecule. $E_{coh}(y)$ is the collective
cohesive energy in dense atomic hydrogen as a function of the
dimensionless Wigner-Seitz cell radius in Bohr units
$y=r_{c}/a_{0}$, calculated in \cite{Khomkin}, where
$r_{c}=(3/4\pi n_{a})^{1/2}$, $n_{a}$ is the density of number of
atoms. For hydrogen and deuterium, these values coincide.
$\eta=\frac{4}{3}\pi[\frac{N_{a}}{V}r_{a}^{3}+\frac{N_{m}}{V}r_{m}^{3}]$
is the total packing parameter expressed in terms of atomic and
molecular radii $r_{a}$ and $r_{m}$, respectively. The first three
terms in (1) describe the mixture of atoms and molecules in
Carnahan-Starling approximation for amounts of molecules and atoms
of the hard core mixture. The fourth term describes the cohesive
energy.

Using the known thermodynamic relations, it is possible to derive
from (1) the expressions for pressure $P$, internal energy $E$,
and chemical potentials of atoms $\mu_{a}$ and molecules $\mu_{m}$
\cite{Khomkin}. The relation $\mu_{m}=2\mu_{a}$ enables derivation
of the equation of dissociation equilibrium \cite{Khomkin}.

\section{III. DPT binodal}

Figure 1 displays the calculated isotherms of dissociation ratio
$\gamma=n_{a}/n$, $n=n_{a}+2n_{m}$, $n_{m}$ is the density of
number of molecules, as functions of density $\rho$ for
temperatures $T = 12000$, $10000$, and $8000$\,K.

At low densities (atomic gas), the dissociative equilibrium curve
tends to unity. As density increases, the value of $\gamma$
decreases and formation of molecules appears. Then, very rapidly,
the dissociation with formation of liquid atomic phase occurs,
which is the DPT.

\begin{figure}
\resizebox{0.4\textwidth}{!}{
\includegraphics{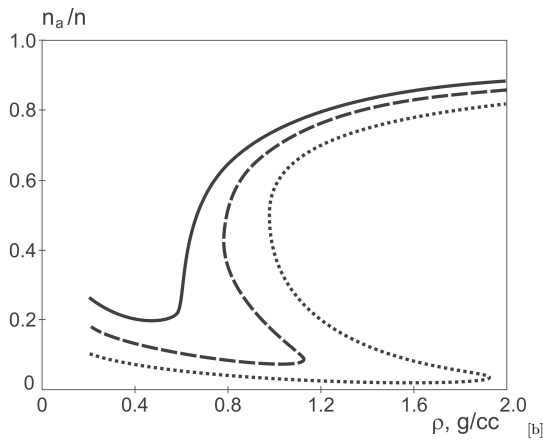}[b]
}\caption{The dissociation equilibrium isotherms: solid, dashed
and dotted curves correspond to T = 12000, 10000 and 8000~K,
respectively.}
\end{figure}

\begin{figure}
\resizebox{0.4\textwidth}{!}{
\includegraphics{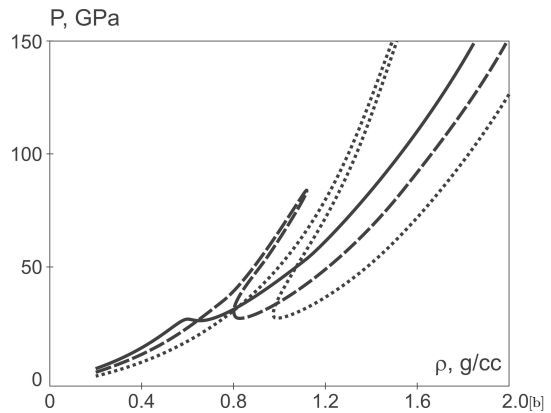}[b]
}\caption{Deuterium pressure dependence on density on isotherm
curves. Curves notations are the same that in Fig.~1.}
\end{figure}

\begin{figure}
\resizebox{0.45\textwidth}{!}{
\includegraphics{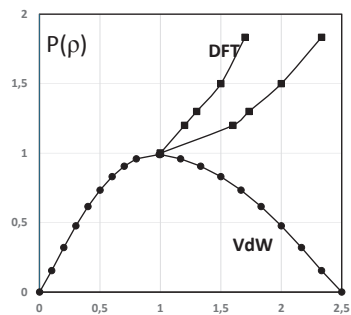}[b]
}\caption{Comparison of DPT and van der Waals (VdW) binodals in
terms of the dimensionless coordinates pressure $\widetilde{P}$-
density $\widetilde{\rho}$.}
\end{figure}

Figure 2 shows the isotherms for the same values of temperature
that in Fig.~1. The curves clearly demonstrate presence of the
unusually shaped van der Waals' loops. We calculated the DPT
binodal using the Maxwell's equal area rule. The following table
presents the calculation results, where $\rho_{g}$ and $\rho_{l}$
are densities of gas and liquid components, respectively.

\begin{table}[b]
\caption{Parameters of DPT binodal}
\begin{ruledtabular}
\begin{tabular}{cccc}
T (K) & P (GPa) & $\rho_{g}$ (g/cm$^3$) & $\rho_{l}$ (g/cm$^3$) \\
\hline
12000 & 30 & 0.6 & 0.6 \\
10500 & 36 & 0.72 & 0.96 \\
10000 & 39 & 0.78 & 1.04 \\
9000 & 45 & 0.9 & 1.2 \\
8000 & 55 & 1.02 & 1.4 \\
\end{tabular}
\end{ruledtabular}
\end{table}

Figure 3 displays the DPT binodal in terms of the dimensionless
units $\widetilde{P}$ and $\widetilde{\rho}$ defined as the ratios
of each thermodynamic quantity to its critical value. Such
presentation facilitates a comparison with the van der Waals (VdW)
binodal. The figure clearly visualizes the difference between two
phase transitions under consideration.

The delocalization degree of the bound electron in the critical
point is close to $1/3$ while the compression delocalization
degree tends to unity (see more details in \cite{Khomkin}).  The
metallization occurs owing to the existence of the conductivity
electrons and to the collective binding energy.

\section{IV. Hugoniot curve and discussion}

The initial conditions in the solution of Rankine-Hugoniot
equation for compression adiabatic curve are $\rho_{0}=0.171$~
g/cm$^3$, $P_{0}=1$~MPa. The DPT for free energy (1) shows certain
peculiarities on the adiabatic curve. The Rankine-Hugoniot
equation in the temperature interval $8000 - 10000$\,K has three
solutions for density.  The first and the second solutions are
stable, while the third one is unstable.

Figure 4 shows the solutions for Hugoniot adiabatic curve where
the molecular and the atomic radii are taken from \cite{Redmer}
and \cite{Khomkin}, respectively.

\begin{figure}
\resizebox{0.5\textwidth}{!}{
\includegraphics{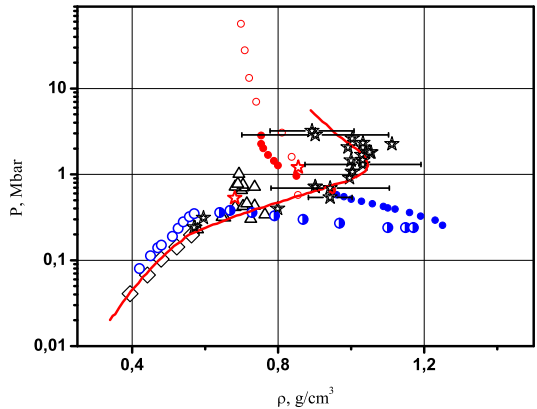}
}\caption{Hugoniot adiabatic curve for deuterium for the soft
molecule diameter $d_{m}=f(n,T)$. Open, solid, and semi-open blue
circles correspond to our first, second, and third solutions,
respectively. Experimental results: black stars are results from
\cite{DaSilva}, red stars are results from \cite{Trunin},
triangles are results from \cite{Knudson}, and diamonds are
results from \cite{Nellis}. Theoretical predictions: solid line
represents paper \cite{Ross}, solid circles represent cell model
\cite{Vorobiev}, and the open circles represent model
\cite{Levashov}.}
\end{figure}

The curves and scatters in Fig.~4 correspond to the experimental
data, our results, and results of the most adequate models of
other authors. The latter include the Ross model \cite{Ross}.
Also, we show the results of calculations in a cell approximation
for the system of electrons and ions \cite{Vorobiev}, and the
results of the Monte-Carlo calculations \cite{Levashov} for the
same system. These results confirm our conclusion on molecular
deuterium metallization under compression. The main qualitative
result is the appearance of the area with ill-defined density.
This unusual phenomenon provides the new interpretation of the
experimental results for deuterium compression.  The discrepancy
of the results of this experiment and results obtained in other
works \cite{Knudson,Nellis,Trunin} appears in result of the
existence of an area with undetermined density (like the two-phase
liquid-vapor state) rather than owing to the uncertainties of the
first experiment \cite{DaSilva}. The anomaly area revealed by our
calculations corresponds to somewhat lower pressure than the
experimental observations. However, we consider the presence of
anomaly area in our model as the most important result of our
work.

\section{V. Conclusions}

In the frame of our model, we found the DPT binodal for dense
dissociating molecular hydrogen.  The DPT binodal is significantly
different from traditional binodal of gas-liquid phase transition
described by the van der Waals equation.  The Hugoniot adiabatic
curve calculated in the frame of our model has an anomaly in the
area of the most scattered experimental data.  We suggested a new
interpretation of the experimental data
\cite{DaSilva,Knudson,Nellis,Trunin}. The anomaly area appears
owing to the presence of two coexisting phases: a molecular fluid
(gas with a density of a liquid) and an atomic metallized liquid.
Thus, density is an ill-defined quantity in this area.

\bibliography{apsKhomkin.bib}

\end{document}